\documentclass[pra,aps,twocolumn]{revtex4}
\usepackage{graphicx}
\usepackage{color}
\usepackage{amsmath}

\begin{document}

\title{Scalable Quantum Computing based on Spin Qubits in CNT QD}

\author{Magdalena Stobi\'nska} \email{magda.stobinska@mpl.mpg.de}
\affiliation{ Institute for Optics, Information and Photonics,
  Erlangen-N\"urnberg University, 91058 Erlangen, Germany,\\ and Max
  Planck Institute for the Science of Light, 91058 Erlangen, Germany}

\author{Gerard J. Milburn} \affiliation{Centre for Quantum Computer
  Technology and School of Physical Sciences, The University of
  Queensland, St Lucia, Queensland 4072, Australia}

\author{Leszek Stobi\'nski} \affiliation{Institute of Physical
  Chemistry, Polish Academy of Sciences, Kasprzaka 44/52, 01-224
  Warsaw, Poland, \\ and Faculty of Materials Science and Engineering,
  Warsaw University of Technology, Wo\l{}oska 141, 02-507 Warsaw, Poland}

\date{\today}

\begin{abstract}
We study experimentally demonstrated single-electron ${}^{12}$C CNT QD
with significant spin-orbit interaction as a scalable quantum computer
candidate. Both electron spin and orbital angular momentum can serve
as a logical qubit for quantum processing. We introduce macroscopic
quantum memory for the system in a form of injected either magnetic or
spin carrying atomic ensemble into the nanotube. CNT provides with a
stable atomic trap in finite temperature and with one-dimensional
nuclear spin lattice in an external magnetic field.  The electron is
coupled to the atomic ensemble through either magnetic or hyperfine
interaction. Easy electron and nuclear spin read-out procedure for
this system is possible.
\end{abstract}

\maketitle

\section{Introduction}

The spin degrees of freedom are promising stationary qubit candidates
useful for quantum information processing, quantum computation and
quantum memory
\cite{Suter2002,Engel2004,Lukin2005,Lukin2007,Morton2008}.  Ease in
accessibility and possibility of coupling to light
\cite{Lovett2009,Schwager2008} makes the electronic spin suitable for
quantum gating. Due to long coherence times of order of
seconds~\cite{Morton2008}, the nuclear qubit is likely to perform as a
quantum memory.

Since the two seminal papers on solid state based quantum computer by
Kane \cite{Kane1998} and Loss and DiVincenzo \cite{Loss1998}, the
theoretical and experimental effort has been pursued to realize a
physical system which will unify the electronic and nuclear spins and
where the spin - spin interactions will be
controllable~\cite{Bose2003,Lovett2008-a}.  This is difficult because
the interaction responsible for coupling, and thus information
transfer between the electronic and nuclear spins, should be strong
enough in order to perform quantum operations effectively. However, it
is not possible to turn the interaction off after the operation is
completed and thus the coupling will result in rapid spin relaxation
and decoherence~\cite{Johnson2005}.  Also unavoidable random nuclear
field fluctuations influence the electron~\cite{Koppens2002}.

Carbon nanotube quantum dots (CNT QD) constitute an excellent physical
system where mutual interactions between one-dimensional spin
structures towards applications in quantum computation can be studied.
Recently it was theoretically shown that relatively weak hyperfine
interactions in ${}^{13}$C (nuclear spin $I=1/2$) enriched CNT QD lead
to the nuclear and electronic self-stabilizing spin ordered phases in
finite temperatures \cite{Loss2009-2} while keeping the electron
decoherence time of order of
microseconds~\cite{Loss2009-1}. Experimentally, in ${}^{13}$C CNT
double QDs the influence of controllable nuclear spin environment on
few-electron spins state, the strength of hyperfine interaction as
well as relaxation and dephasing times were
determined~\cite{Marcus2009-1,Marcus2009-2}.

In this paper we study recently experimentally demonstrated
single-electron ${}^{12}$C CNT QD with significant spin-orbit
interaction \cite{Ilani2008} as a scalable quantum computer candidate
obeying all the DiVincenzo requirements~\cite{DiVincenzo2000}. In this
system both electron spin and orbital angular momentum can serve as a
qubit for processing. Anomalous spin relaxation times for this system
were found \cite{Bulaev2008}. Since ${}^{12}$C atom does not carry
nuclear spin, we provide macroscopic quantum memory for the system in
a form of injected either magnetic or spin carrying atomic ensemble
into the nanotube.  Thus, the electron is coupled to the atomic
ensemble through either hyperfine or magnetic interaction. After
information recording onto the atomic ensemble the electron can be off
loaded to turn the electron-nuclear interactions off. Easy electron
spin read-out procedure for this system is possible.

Originally, encapsulation of molecules and other nano-objects in CNT
was extensively studied towards high energy storage and its
controllable realise (see e.g. \cite{H-CNT, N-CNT}).  However, up to
our best knowledge it has never been investigated towards a neutral
atomic trap and macroscopic quantum memory applications.  Carbon
nanotube provides with stable atomic trap in finite temperature
(preventing from possible molecular phase formulation) as well as with
one-dimensional nuclear spin lattice in an external magnetic field.

Similar idea to our was explored for intercalated fullerens. Single
nitrogen (${}^{14}$N, ${}^{15}$N) or phosphorus (${}^{31}$P) atoms
injected into a $C_{60}$ molecule would serve as a quantum memory
\cite{Harneit2002,Feng2004,Ju2007}. However, for these schemes the
read-out protocol turned out to be very challenging. Moreover, our
scheme involves macroscopic atomic ensemble, thus enhancing the
efficiency of interaction with a single electron and enabling coupling
to light and memory read-out via Faraday effect.

This paper is organized as follows. In section \ref{sec1} we briefly
identify the energy eigenstates of the ${}^{12}$C CNT QD and possible
logical qubit realizations. Next, we discuss the implementations of
single- and two-qubits rotations necessary for gating. We discuss the
idea of atomic trap for one-dimensional nuclear spin lattice inside
the CNT in section \ref{sec3} and continue towards its applications as
a quantum memory. Finally, we conclude.

\section{Scalable QC candidate} \label{sec1}

\begin{figure}
\begin{center}
   \scalebox{0.3}{\includegraphics{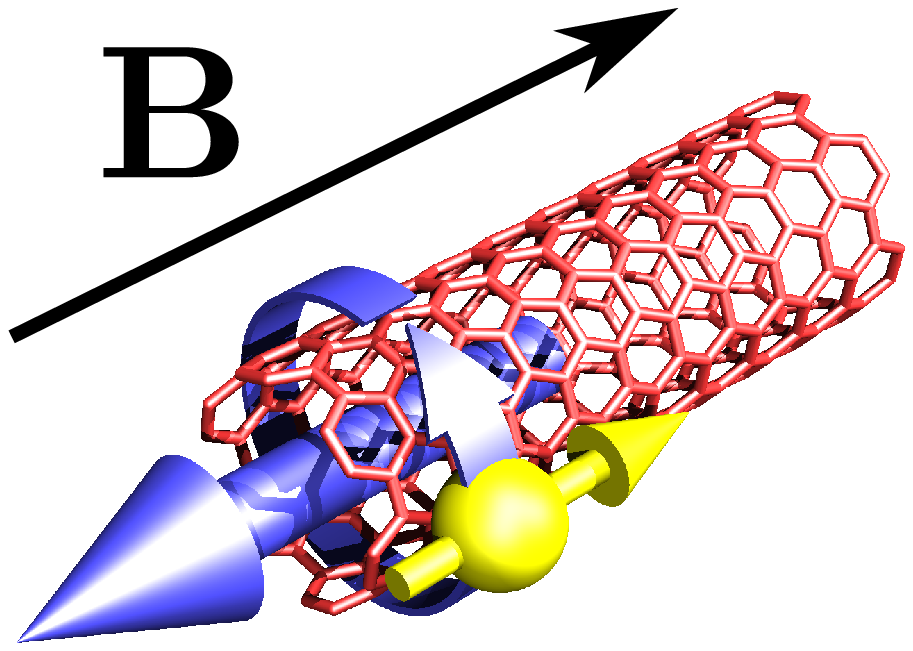}}
   \scalebox{0.3}{\includegraphics{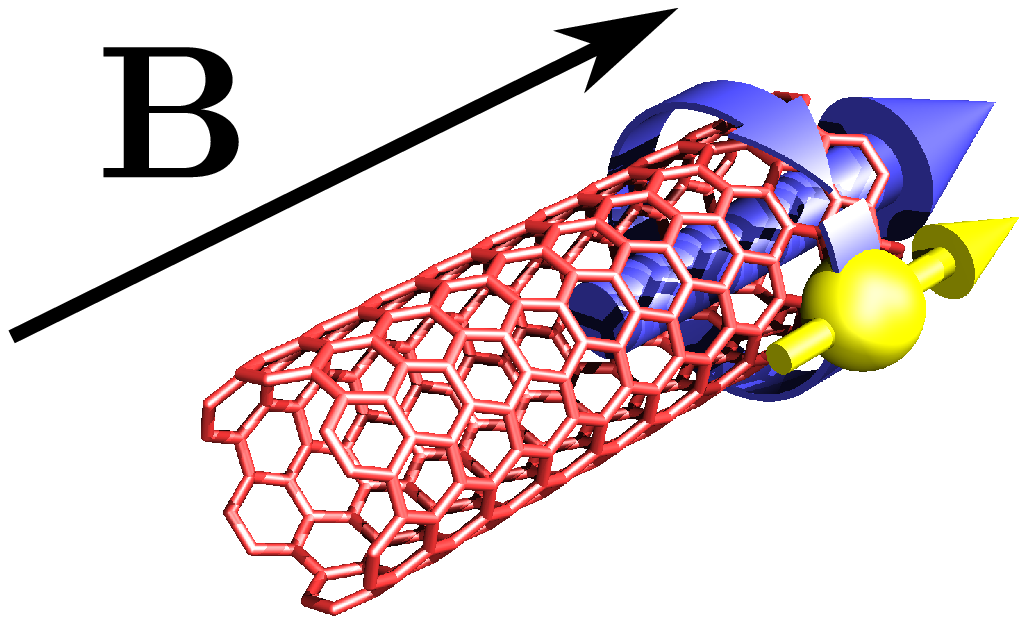}}
\end{center}
\caption{A single-electron CNT QD in external magnetic field. Spin
  (yellow arrow) and orbital angular momentum (blue arrow) can be
  parallel or antiparallel.}
\label{fig1}
\end{figure}

\begin{figure}
\begin{center}
   \scalebox{0.4}{\includegraphics{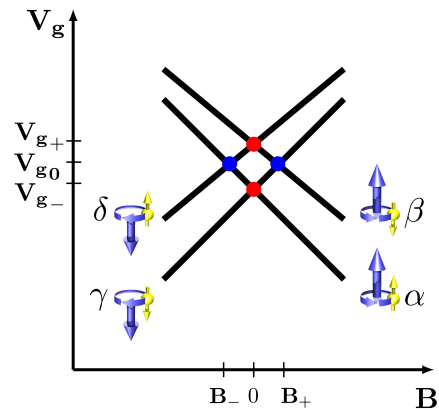}}
\end{center}
\caption{Electronic states in CNT QD in presence of an external
  magnetic field.}
\label{fig2}
\end{figure}

Recently, a single-electron carbon nanotube quantum dot (CNT QD) with
a significant spin-orbit interaction in regime of mK temperatures has
been reported \cite{Ilani2008}. The electron was loaded onto a
small-band gap 'quasi-metallic' single wall nanotube (SWNT) and an
external magnetic field $B$ along the nanotube (zed direction) was
applied. The field constitutes a quantization axis for the electron's
spin and orbital angular momentum degrees of freedom, see
Fig. \ref{fig1}.
The spin $\hat{S}_z$ projection is oriented either parallely
$|1/2\rangle$ or anti-parallely $|-1/2\rangle$ to the field direction.
The electron circumvents clockwise $|\uparrow\rangle$ or
anti-clockwise $|\downarrow\rangle$ the CNT.  The measured energy
spectrum versus the magnetic field and a gate voltage $V_g$ shows four
possible system quantum states ($\alpha$, $\gamma$, $\beta$,
$\delta$). They become pairwise energetically degenerate for some
specific values of the external parameters, see Fig. \ref{fig2}.
The states are two-dimensional product states $\alpha= |1/2\rangle_S
|\uparrow\rangle_L$, $\beta = |-1/2\rangle_S |\uparrow\rangle_L$,
$\gamma=|-1/2\rangle_S |\downarrow\rangle_L$, $\delta = |1/2\rangle_S
|\downarrow\rangle_L$, where $S$ and $L$ stand for spin and angular
momentum Hilbert space respectively. For $B=0$ and $V_g=V_{g-}$
($V_g=V_{g+}$) the states $\alpha$ and $\gamma$ ($\beta$ and $\delta$)
acquire the same energy. Therefore, the system is naturally described
by their superpositions, the Kramers doublets being the maximally
entangled Bell states at the same time
\begin{eqnarray}
|\Omega_1\rangle &=& |1/2\rangle_S |\uparrow\rangle_L +
e^{i\varphi_1} |-1/2\rangle_S
|\downarrow\rangle_L,
\label{K1}\\ |\Omega_2\rangle &=& |-1/2\rangle_S
|\uparrow\rangle_L + e^{i\varphi_2} |1/2\rangle_S
|\downarrow\rangle_L,\label{K2}
\end{eqnarray}
where $\varphi_i$ are relative phases.  The energy splitting between
them results from the spin-orbit coupling and is equal to $\Delta_{SO}
= 0.4$~\cite{Ilani2008}. For $B=B_+$ ($B=B_-$) and $V_g=V_{g0}$ the
states $\alpha$ and $\delta$ ($\gamma$ and $\beta$) are almost
degenerate with a small splitting $\Delta_{KK'}=65\mu$eV emerging from
the Dirac cones anti-mixing. Nevertheless, is it possible to couple
$\gamma$ and $\beta$ ($\alpha$ and $\delta$) in a non-resonant
transition and describe the system by a product state of the form
\begin{eqnarray}
|\omega_1\rangle &=& |-1/2\rangle_S \, (|\downarrow\rangle_L +
e^{i\varphi_3} |\uparrow\rangle_L),\\ |\omega_2\rangle &=&
|1/2\rangle_S \, (|\uparrow\rangle_L + e^{i\varphi_4}
|\downarrow\rangle_L).
\end{eqnarray}
please note, that angular momentum part of the state is given by a its
eigenstates superposition.

Both electron's degrees of freedom, spin and angular momentum
$\hat{L}_z$ (valley degree of freedom), can be used for encoding
logical qubits with experimentally feasible one- and two-qubit
rotations since they are easily accessible for the external operations.
Tunneling spectroscopy realized by rising and lowering the Fermi level
provides a read-out scheme for spin state \cite{Ilani2008}.

\subsection{System Hamiltonian}  

Due to barrier potential in CNT QD electron's longitudinal momentum
component $k_{\parallel}$ is quantized. The nanotube periodic boundary
conditions enforce quantization of perpendicular $k_{\perp}$ component
as well. Assuming the lowest electron momentum excitation and
$k_{\parallel}=0$ what corresponds to the experimental
situation~\cite{Ilani2008} (for a metallic zig-zag CNT) we identify the
orbital angular momentum quantum number to be $L=1$. The eigenstates
of $\hat{L}_z$ component correspond to its eigenvalues $m_l$ in the
following way $|\uparrow\rangle_L = |m_l=1\rangle_L$,
$|\downarrow\rangle_L = |m_l=-1\rangle_L$, where $m_l \le |L|$. Next,
we introduce the total angular momentum quantum number
$J=1/2,3/2$. This values of $J$ correspond to sp hybridization
\cite{Loss2009-1}. 

Since the smallest diameter of a stable carbon nanotube is
$0.6-0.7$~nm, the system can be simulated as a hydrogen-like structure
with an electron being 14 times further from the nucleus or as a
Rydberg-like atom with an electron on a nanoshell. In this case a
Hamiltonian of the system reads
\begin{equation}
 H = \frac{1}{2m^2c^2}\,\frac{1}{r}\,\frac{dV(r)}{dr}
 \hat{L}_z\cdot\hat{S}_z - \frac{e B}{2 m} \left(\hat{L}_z +
 2\hat{S}_z\right).
\end{equation}
The states $\alpha$ and $\gamma$ are the energy eigenstates
\begin{eqnarray}
\alpha \!\!&=&\!\! |J\!\!=\!\!3/2, m_j\!\!=\!\!3/2\rangle,
\nonumber\\ \gamma \!\!&=& \!\!|J\!\!=\!\!3/2,
m_j\!\!=\!\!-3/2\rangle,\nonumber
\end{eqnarray}
whereas the other two are not
\begin{eqnarray}
\delta \!\!&=&\!\! \left(|J\!\!=\!\!3/2, m_j\!\!=\!\!-1/2\rangle -
\sqrt{2} |J\!\!=\!\!1/2, m_j\!\!=\!\!-1/2\rangle\right)/\sqrt{3},
\nonumber\\ \beta \!\!&=&\!\!  \left(|J\!\!=\!\!3/2,
m_j\!\!=\!\!1/2\rangle + \sqrt{2} |J\!\!=\!\!1/2,
m_j\!\!=\!\!1/2\rangle\right)/\sqrt{3}\nonumber.
\end{eqnarray}

\subsection{Logical qubit in $\vec{B}=0$ regime}

Let us define the logical qubit as the energy eigenstates $|1\rangle =
\alpha$ and $|0\rangle = \gamma$. One-qubit rotation requires creating
superposition of the form $\alpha + e^{i \varphi_1} \gamma$ for an
arbitrary phase value $\varphi_1$. This is possible by applying
magnetic field kick. For $\vec{B}!=0$ the states $\alpha$ and $\gamma$
correspond to different energy eigenstates $E_{\alpha}$ and
$E_{\gamma}$ and therefore evolve with different phases $\alpha + e^{i
  \varphi_1} e^{-i (E_{\alpha} - E_{\gamma}) t/\hbar}\gamma$.

\subsection{Logical qubit in $\vec{B}=B_{+}$ regime}

Let us encode the logical qubit in orbital angular momentum $|1\rangle
= \gamma$ and $|0\rangle =\beta$. Since the states are not fully
energetically degenerate, one-qubit rotations have to induced by a
near resonant transition.  Simple algebra $\gamma = \frac{1}{2}(\gamma
+ \beta) + \frac{1}{2}(\gamma - \beta)$ shows that state $\gamma$
addressed by a microwave pulse with $\nu=65\kern.25em\mathrm{GHz}$
evolves to
\begin{equation}
  \frac{1}{2}(\gamma + \beta) + \frac{1}{2}e^{i \Delta t}(\gamma -
  \beta) = \cos(\Delta t)\gamma + \sin(\Delta t)\beta.
\end{equation}

\subsection{Coupling between two CNT QDs}

\begin{figure}[t]
\begin{center}
   \scalebox{0.2}{\includegraphics{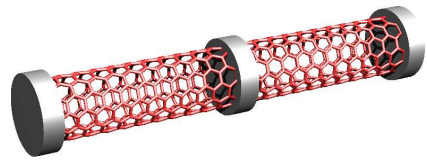}}
\end{center}
\caption{Two-qubit coupling architecture for single-electron CNT QDs.}
\label{fig5}
\end{figure}

Two-qubit rotations can be realized using either magnetic dipole
interaction
\begin{equation}
H = - \frac{\mu_0}{2\pi}\,\frac{1}{r^3}
(\hat{L}_{z1}\cdot\hat{L}_{z2})
\end{equation}
present when the two CNT are located one after another (see
Fig. \ref{fig5}) or a Heisenberg spin-spin exchange interaction when
two electrons are brought close to each other.

\section{1D nuclear spin lattice trap}  \label{sec3}

A numerical analysis of a potential energy for a system consisting of
a single empty SWCNT and gas of hydrogen, nitrogen and phosphorus atoms
in $T\!=\!0$K and in finite temperature $T\!>\!0$K was performed.  The
results showed self-induced intercalation of the atoms into the
nanotube due to the van der Waals interactions. We employed HyperChem
program and performed our calculations using Molecular Mechanics MM+
method for $T\!=\!0$K and Molecular Dynamics with step size
$10^{-4}$ps and runtime $1$ps for $0\!<\!T\!<\!100$K.  Due to the
symmetric and periodic structure of the nanotube, the atoms form a
string located on the nanotube axis with almost equal spacing between
them of approximately $3\mathrm{\AA}$ for hydrogen atoms and $3.5
\mathrm{\AA}$ nitrogen atoms. Moreover, the trapping energy is strong
enough to prevent from forming molecular phase of the H and N
atoms. Thus, the nanotube provides with a stable atomic trap. If the
atoms carry spin, in an external magnetic field they form a 1D nuclear
spin lattice. The calculations were obtained for the following
nanotubes with similar diameters (around $0.7$nm): CNT(5,5)
(armchair), CNT(9,0) (zig-zag), CNT(8,2) (chiral). The CNT were filled
with 8 atoms. In order to avoid the edge effects CNTs were longer than
chain of atoms. The depth of the trapping potential at all atomic
positions was the same since the atoms when slightly increasing
temperature acquired the same thermal oscillations frequency remained
still trapped.  Any off-axial position of the chain was self-corrected
into the axial location.

Our results remain in agreement with other findings reported in
literature. For example, the stability conditions of the carbon chains
inside CNT as a function of CNT diameter were studied in
\cite{Liu2003}.  The binding energy was found to be of order of
$0.1$eV. The stretching mode frequency for the chain inside CNT was
lower than this of the isolated chain as was of order of
$2000cm^{-1}$.

\bigskip
\begin{figure}[t]
\begin{center}
   \scalebox{0.3}{\includegraphics{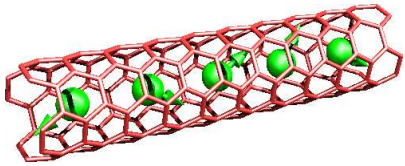}}
\end{center}
\caption{A single wall carbon nanotube provides with a stable atomic
  trap.}
\label{fig6}
\end{figure}

\subsection{Application towards quantum memory}  \label{sec4}

\begin{figure}[h]
\begin{center}
   \scalebox{0.3}{\includegraphics{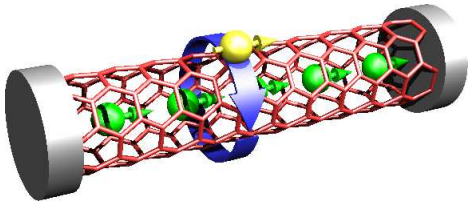}}
\end{center}
\caption{${}^{12}$C CNT QD with injected macroscopic atomic 1D spin
  lattice.}
\label{QC}
\end{figure}

If the spin carrying or magnetic atomic ensemble was injected into the
${}^{12}$C CNT QD, in external magnetic field the system could mimic
quantum computer with long lived quantum memory (see Fig. \ref{QC}).
Since carbon $C^{12}$ has no nuclear spin, the system allows for
exploiting hyperfine interaction between the macroscopic nuclear spin
$\hat{I}$ lattice inside the CNT and the electron
\begin{equation}
H\!=\!-\frac{\mu_0}{4\pi}\,\frac{2\mu_B\mu_n g_p}{\hbar^2}\,
\frac{1}{r^3} \left( \hat{I}\cdot\hat{L}\!+\!3
(\hat{I}\cdot\bar{n})(\hat{S}\cdot\bar{n})\!-\!\hat{I}\cdot\hat{S}
\right).
\end{equation}

Also angular momentum coupling could be realized if the atoms were
magnetic.  Read-out scheme for quantum memory is possible via Faraday
effect.

\section{Conclusions}

We have discussed a single-electron ${}^{12}$C CNT QD as a candidate
for a scalable quantum computer.  This system has already been
realized experimentally. It is possible to define logical qubits for
the electron degrees of freedom and perform their rotations. Nuclear
quantum memory could be added to the system if additionally spin
carrying or magnetic atoms were injected into the CNT. Intercalation
of CNT of any chirality is already a standard procedure within current
technology.

The source of relaxation and decoherence effects for this enriched
system remains an open question. They can result from Coulomb
interactions or coupling between the electron and CNT phonons
\cite{Bulaev2008}.

\section{Acknowledgments}
M.S. thanks to A. Aiello, G. Burkard, P. Horodecki and D. Loss for
stimulating discussions.

\vfil


\begin{thebibliography}{99}

\bibitem{Suter2002} D. Suter and K. Lim, \pra {\bf 65}, 052309 (2002).

\bibitem{Engel2004} H.-A. Engel, L. P. Kouwenhoven, D. Loss, and C. M. Marcus,
  Quantum Information Processing {\bf 3}, 1 (2004).

\bibitem{Lukin2005} J. M. Taylor, H.-A. Engel, W. D\"ur, A. Yacoby,
  C. M. Marcus, P. Zoller, and M. D. Lukin, Nature Physics {\bf 1},
  177 (2005).

\bibitem{Lukin2007} M. V. Gurudev Dutt, L. Childress, L. Jiang,
  E. Togan, J. Maze, F. Jelezko, A. S. Zibrov, P. R. Hemmer,
  M. D. Lukin, Science {\bf 316}, 1312 (2007).

\bibitem{Morton2008} J. J. L. Morton et al, Nature {\bf 455}, 1085
  (2008).

\bibitem{Loss} J. Lehmann, A. Gaita-Arino, E. Coronado, and D. Loss,
  Nature Nanotechnology {\bf 2}, 312 (2007). 

\bibitem{Lovett2009} M. Schaffry et al., arXiv:0911.5320v2. 

\bibitem{Schwager2008} H. Schwager, J. I. Cirac, G. Giedke,
  arXiv:0810.4488v2.

\bibitem{Kane1998} B. E. Kane, Nature {\bf 393}, 133 (1998).

\bibitem{Loss1998} D. Loss and D. P. DiVincenzo, \pra {\bf 57}, 120
  (1998).

\bibitem{Bose2003} S. C. Benjamin, and S. Bose, \prl {\bf 90}, 247901 (2003).

\bibitem{Lovett2008-a} E. M.  Gauger, P. P. Rohde, A. M. Stoneham, and B. W.
  Lovett, New J. Phys. {\bf 10}, 073027 (2008).

\bibitem{Johnson2005} A. C. Johnson, J. R. Petta, J. M. Taylor,
  A. Yacoby, M. D. Lukin, C. M. Marcus, M. P. Hanson, and
  A. C. Gossard, Nature {\bf 435}, 925 (2005).

\bibitem{Koppens2002} F. H. L. Koppens et al., Science {\bf 309}, 1346
  (2002).

\bibitem{Loss2009-2} B. Braunecker, P. Simon, and D. Loss, \prl {\bf
  102}, 116403 (2009).

\bibitem{Loss2009-1} J. Fischer, B. Trauzettel, and D. Loss, \prb {\bf
  80}, 155401 (2009).

\bibitem{Marcus2009-1} H. O. H. Churchill, A. J. Bestwick, J.W. Harlow,
  F. Kuemmeth, D. Marcos, C. H. Stwertka, S. K.Watson, and
  C. M. Marcus, Nature Physics {\bf 5}, 321 (2009).

\bibitem{Marcus2009-2} H. O. H. Churchill, F. Kuemmeth, J. W. Harlow,
  A. J. Bestwick, E. I. Rashba, K. Flensberg, C. H. Stwertka,
  T. Taychatanapat, S. K. Watson, and C. M. Marcus, \prl {\bf 102},
  166802 (2009).

\bibitem{Ilani2008} F. Kummeth, S. Ilani, D. C. Ralph, and
  P. L. McEuen, Nature {\bf 452}, 448 (2008).

\bibitem{DiVincenzo2000} D. P. DiVincenzo, Science {\bf 270}, 255
  (1995).

\bibitem{Bulaev2008} D. V. Bulaev, B. Trauzettel, and D. Loss, \prb
  {\bf 77}, 235301 (2008).

\bibitem{H-CNT} A. C. Dillon, K. M. Jones, T. A. Bekkedahl,
  C. H. Klang, D. S. Bethune, and M. J. Heben, Nature {\bf 386}, 377
  (1997).

\bibitem{N-CNT} H. Abou-Rachid, A. Hu, V. Timoshevskii, Y. Song, and
  L.-S. Lussier, \prl {\bf 100}, 196401 (2008).
 
\bibitem{Harneit2002} W. Harneit, \pra {\bf 65}, 032322 (2002).

\bibitem{Feng2004} M. Feng and J. Twamley, \pra {\bf 70}, 030303(R)
  (2004).

\bibitem{Ju2007} C. Ju, D. Suter, and J. Du, \pra {\bf 75}, 012318
  (2007).

\bibitem{Liu2003} Y. Liu, R. O. Jones, X. Zhao, and Y. Ando, \prb {\bf
  68}, 125413 (2003).

\end{thebibliography}
\end{document}